\documentclass[seceq]{ptptex}
\title{Schwinger's Principle and Gauge Fixing in the Free Electromagnetic
Field }

\author{C. A. M. \textsc{de Melo$^{1,}$\footnote{%
email: cassius@ift.unesp.br}}, B. M. \textsc{Pimentel$^{1,}$\footnote{%
email: pimentel@ift.unesp.br}} and P. J. \textsc{Pomp\'{e}ia$^{1,\,2,}$%
\footnote{%
email: pompeia@ift.unesp.br}}
}

\inst{%         %Affiliation, neglected when [addenda] or [errata]
$^{1}$Instituto de F\'{\i}sica Te\'{o}rica

Universidade Estadual Paulista

Rua Pamplona,145

01405-900 - S\~{a}o Paulo, S.P.

Brazil

\vskip0.3cm

$^{2}$Centro T\'{e}cnico Aeroespacial

Instituto de Fomento e Coordena\c{c}\~{a}o Industrial

Divis\~{a}o de Confiabilidade Metrol\'{o}gica

Pra\c{c}a Marechal Eduardo Gomes, 50

12228-901 - S\~{a}o Jos\'{e} dos Campos, S.P.

Brazil }

\abst{%         %this abstract is neglected when [addenda] or [errata]
 A manifestly covariant treatment of the  free quantum eletromagnetic field, in a linear
 covariant gauge, is implemented employing the Schwinger's Variational Principle and
 the B-field formalism. It is also discussed the abelian Proca's model as an example of
 a system without constraints.
}

\begin{document}

\maketitle

\section{Introduction}

The covariant quantization of the electromagnetic field is one of
the most peculiar problems of Quantum Field Theory because of the
masslessness of the photon. In spite of its vectorial nature, only
the two transverse components of the photon are observable, and the
third freedom yields the Coulomb interaction between charged
particles. The first consistent\ manifestly covariant quantization
of the electromagnetic field was formulated by Gupta \cite{Gupta}
and independently by Bleuler \cite{Bleuler}, describing the photon
in the Fermi gauge and the Lorentz condition not being an operator
identity but a restriction which is imposed on the physical states.
As it is well known, in their theory, the state vectors do not
necessarily have\ positive norm, and the space spanned by them is an
indefinite metric Hilbert space. It is often convenient
theoretically to use an gauge independent quantization, however,
such a method is not known nowadays. Concerning to covariance,
Schwinger has proposed a general principle that describes the
Quantum Field Theory in a manifestly covariant way.

Schwinger's Variational Principle was implemented in 1951
\cite{QuanField} as a powerful method to extract the kinematical
(commutation relations) and dynamical (equations of motion)
qualities from only one principle in a self-consistent and covariant
manner. With this method Schwinger wished to systematize the Quantum
Electrodynamics (QED$_{4}$) in such a way that the various
developments and results could be obtained from a theory of the
fundamental processes involved.

On the other hand, QED$_{4}$ is a Gauge Theory with non-trivial
dynamic constraints among its degrees of freedom, and therefore it
is necessary a careful analysis in the establishment of its degrees
of freedom and the consequences of this symmetry. A particular and
profitable way to broach\ the freedom of gauge of the free
electromagnetic field was stablished by Nakanishi in 1966-67
\cite{NakaArt} for covariant linear gauges, through an auxiliary\
scalar field $B$. In a serie of subsequent works, Nakanishi and
colaborators applied this formalism to analyse several systems as
General Relativity \cite{NakGrav1}, Proca's field \cite{NakaProca}
and the Model of Higgs \cite{BfieldHiggs2}.

In this paper, we intend to do a brief analysis of the free
eletromagnetic field employing Schwinger's Variational Principle,
using Palatini's variation method \cite{Palatini}\ , and the
formalism of Nakanishi's auxiliary field to treat the freedom of
gauge of the system. The Palatini's method of variation was
originally developed in the context of the General Relativity Theory
as a manner of finding a coupled system of first order equations for
the gravitational potential $g_{\mu \nu }$ and its ``momentum''\
$\Gamma _{\mu \nu }^{\,\,\rho }$.

In section \ref{sec1} we will give a brief description of
Schwinger's Principle. Next, in section \ref{sec2} we make an
application to free eletromagnetic field using the formalism of
auxiliary field for covariant linear gauges, obtaining the equations
of motion and commutation relations. To make a comparison, we make
the same analysis to the vectorial massive field in section
\ref{sec3}. To determine the dynamical degrees of freedom
of the system, we analyse the Cauchy Problem of the both systems in section %
\ref{sec4}. At last, we make some comments, showing our conclusions.

\section{Schwinger's Variational Principle\label{sec1}}

In Schwinger's approach, the problem of the relativistic quantum
dynamic is to determine the transformation function $\left\langle
a^{2},\sigma ^{2}\right| \left| a^{1},\sigma ^{1}\right\rangle $,
where $\sigma ^{1}$\
and $\sigma ^{2}$\ are two distinct space-like surfaces, and $a^{1}$\ and $%
a^{2}$\ are the eingenvalues of a complete set $A$\ that describe
the system, associated to each surface respectively \cite{Legacy}.

The fundamental issue of this formulation of Quantum Field Theory is
the
Schwinger's Variational Principle, that can be written as%
\[
\delta \left\langle a^{2},\sigma ^{2}\right| \left| a^{1},\sigma
^{1}\right\rangle =i\left\langle a^{2},\sigma ^{2}\right| \delta
S\left| a^{1},\sigma ^{1}\right\rangle =i\left\langle a^{2},\sigma
^{2}\right| \delta \left[ \int_{\sigma ^{1}}^{\sigma ^{2}}d\Omega
L\left( \chi _{a}\left( x\right) ,x\right) \right] \left|
a^{1},\sigma ^{1}\right\rangle =
\]%
\[
=i\left\langle a^{2},\sigma ^{2}\right| \left( \int_{\sigma
^{2}}-\int_{\sigma ^{1}}\right) d\sigma _{\mu }G^{\mu }\left( \sigma
\right) \left| a^{1},\sigma ^{1}\right\rangle \,,
\]%
where $L$ is the Lagrangian density operator that describe the
system of interest and $G^{\mu }\left( \sigma \right) $\ is the
infinitesimal generator of general variations on the states. This
generator may contain information on kinematical and dynamical
aspects of the system, farther on symmetries. Moreover, in virtue of
the invariance of the structure of the
Variational Principle under an addition of a total divergence,%
\begin{eqnarray*}
L &\rightarrow &\bar{L}=L+\partial _{\mu }\Lambda ^{\mu } \\
G^{\mu } &\rightarrow &\bar{G}^{\mu }=G^{\mu }+\Lambda ^{\mu }
\end{eqnarray*}%
we can change the generator and therefore choose the most convenient
representation for the calculations\footnote{%
Naturally this choice alters our interpretation of the generator,
and it
results in different functional relations among the\ fundamental operators.}%
. In practical employments of the above expression, it maybe
necessary the use of boundary conditions for the fields on the
limits of space-like surfaces, or, equivalently, on a time-like
surface connecting both $\sigma $ surfaces, that delimit the
space-time volume of interest. This principle is the starting point
to several developments, such as: commutation relations, equations
of motion, perturbation theory, construction of propagators,
scattering, etc. Moreover, it can be shown that the canonical
commutation relations obtained from the variational principle are
covariant, and also
give us the infinitesimal generators of Lorentz transformations \cite%
{SchEletro1, QuanField}.

\section{B-Field Formalism and Free Electromagnetic Field\label{sec2}}

To implement the gauge fixing of the free eletromagnetic field in a
covariant manner, we introduce an\ hermitian auxiliary scalar field
$B\left( x\right) $. The introduction of this new field was inspired
in the generalized canonical dynamics proposed by Utiyama
\cite{Utiyama}. Therefore, suppose the following Lagrangian density
operator \cite{NakaLivro}
for the eletromagnetic field with an hermitian auxiliary scalar field, $%
B\left( x\right) $\footnote{%
It is interesting to note that in the his original paper Nakanishi
has proposed a kinetic term proportional to $\left( \partial _{\mu
}B\right) \left( \partial ^{\mu }B\right) $ which had no effect in
the canonical commutation relations and equations of motion, and
therefore no kinematical or dynamical consequences.}:
\[
L=-\frac{1}{2}\left\{ F^{\mu \nu },\frac{1}{2}\left( \partial _{\mu
}A_{\nu }-\partial _{\nu }A_{\mu }\right) \right\}
+\frac{1}{4}F^{\mu \nu }F_{\mu
\nu }+\frac{1}{2}\left\{ B,\partial ^{\mu }A_{\mu }\right\} +\frac{1}{2}%
\alpha B^{2}
\]%
where $\left\{ a,b\right\} $ express the anti-commutator of the
field operators $a$ and $b$; $\alpha $\ \ is a \ fixed scalar
parameter, and it is
implicit the hypothesis that the hermitian field operators $A_{\mu }$\ , $%
F^{\mu \nu }$ e $B$\ are independent, as proposed by Palatini's method \cite%
{Palatini}. The field operators are assumed to be analytical
applications of ${\cal M}^{4}\equiv {\Bbb R}^{\left( 3,1\right) }$
(where $x^{\mu }$ is not quantized) in the functional space of
operators, in such a way that the raising and lowering of indexes
are made by the non-quantized Minkowski's metric $\eta ^{\mu \nu }$
with signature $\left( +,-,-,-\right) $.

This Lagrangian density operator is gauge invariant since%
\[
F^{\mu \nu }\rightarrow F^{\mu \nu }\,,\,B\rightarrow B\,,\,A_{\mu
}\rightarrow A_{\mu }+\partial _{\mu }\Phi
\]%
\[
L\rightarrow L+\square \Phi
\]%
where $\partial _{\mu }\partial ^{\mu }\Phi =\square \Phi \equiv 0$ and $%
\Phi \left( x\right) $\ is a scalar field operator.

If we consider only variations at fixed point, we have%
\begin{align*}
\delta L& =\frac{1}{2}\partial _{\mu }\left\{ F^{\left[ \nu \mu
\right] }+B\eta ^{\nu \mu },\delta A_{\nu }\right\}
-\frac{1}{4}\left\{ \delta F^{\mu \nu },\left( \partial _{\mu
}A_{\nu }-\partial _{\nu }A_{\mu }\right)
-F_{\mu \nu }\right\} + \\
& +\frac{1}{2}\left\{ \partial _{\mu }F^{\left[ \mu \nu \right]
}-\partial ^{\mu }B,\delta A_{\nu }\right\} +\frac{1}{2}\left\{
\delta B,\eta ^{\mu \nu }\partial _{\nu }A_{\mu }+\alpha B\right\}
\,,
\end{align*}%
where we use a compact notation, $F^{\left[ \nu \mu \right] }\equiv \frac{1}{%
2}\left( F^{\nu \mu }-F^{\mu \nu }\right) $. Integrating the variation of $%
\delta L$ in a volume $\Omega $ we see that the first term of the
second member can be transformed in a surface's integral with the
support of Gauss's theorem. The other terms give us the equations of
motion for the
fields%
\begin{align}
\partial _{\mu }F^{\left[ \mu \nu \right] }-\partial ^{\nu }B& =0
\label{EqMaxwell} \\
\partial _{\mu }A_{\nu }-\partial _{\nu }A_{\mu }& =F_{\mu \nu }
\label{DefF} \\
\partial ^{\mu }A_{\mu }+\alpha B& =0  \label{Lorentz}
\end{align}

We know that it is exactly the surface term that gives us the
generator of
infinitesimal field transformations, $G$ \cite{QuanField}. It follows%
\[
G_{2}-G_{1}=\left( \int_{\sigma _{1}}-\int_{\sigma _{2}}\right)
d\sigma _{\mu }^{\bar{x}}\left[ \frac{1}{2}\left\{ F^{\left[ \nu \mu
\right] }\left( \bar{x}\right) +B\left( \bar{x}\right) \eta ^{\nu
\mu },\delta A_{\nu }\left( \bar{x}\right) \right\} \right] \,,
\]%
here we used the boundary condition of null field on the spatial
infinity. Then, applying the generator of the functional variations
over one of the surfaces, we found that the induced variation on the
$A_{\lambda }$ operator
is%
\[
\delta A_{\lambda }\left( x\right) =-i\left[ A_{\lambda }\left( x\right) ,G%
\right] \,,
\]%
where $\left( x-\bar{x}\right) ^{2}<0$.

With the supposition that $\delta A_{\nu }$ is proportional to
identity,
which means that $\delta A_{\nu }$ commutes with all other operators, we have%
\begin{equation}
\delta A_{\lambda }\left( x\right) =-i\int_{\sigma }d\sigma _{\mu }^{\bar{x}}%
\left[ A_{\lambda }\left( x\right) ,\pi ^{\nu \mu }\left( \bar{x}\right) %
\right] \delta A_{\nu }\left( \bar{x}\right) \,,  \label{varA}
\end{equation}%
where $\pi ^{\nu \mu }\left( \bar{x}\right) =F^{\left[ \nu \mu
\right] }\left( \bar{x}\right) +B\left( \bar{x}\right) \eta ^{\nu
\mu }$.

Making a choice of frame in such a way that $x^{0}=\bar{x}^{0}$, we
see that the only possible solution for these equations is that the
fields satisfy,
over the surface $\sigma _{0}$, the identities%
\[
\left[ A_{\lambda }\left( {\bf x}\right) ,\left( F^{\left[ \nu
0\right] }\left( {\bf \bar{x}}\right) +\eta ^{\nu 0}B\left( {\bf
\bar{x}}\right)
\right) \right] _{0}=i\delta _{\lambda }^{\,\,\nu }\delta \left( {\bf x-\bar{%
x}}\right) \,,
\]%
the suffix $0$ in the commutator indicates that the involved
operators are both considered at the same instant of time, fixed by
the choice of a space-like hypersurface. This expression show us
that the canonical momentum conjugated to $A_{\lambda }$ must be
$F^{\left[ \nu 0\right] }+\eta ^{\nu 0}B $.

Employing the generator of the functional variations we find the
induced variation on the $F^{\alpha\beta}$ and $B$ operators.We can
still consider
the independence of symmetrical and antisymmetrical parts of $%
F^{\alpha\beta} $, and see that%
\begin{equation}
\delta\pi^{\alpha\beta}\left( x\right) =-i\int_{\sigma}d\sigma_{\mu}^{\bar{x}%
}\left[ \pi^{\alpha\beta}\left( x\right) ,\pi^{\nu\mu}\left( \bar{x}\right) %
\right] \delta A_{\nu}\left( \bar{x}\right) \,.  \label{varPi}
\end{equation}

On the other hand, adding a surface term we have the following
generator of
functional variations:%
\[
\bar{G}=-\int_{\sigma}d\sigma_{\mu}\frac{1}{2}\left\{ \delta F^{\left[ \nu\mu%
\right] }+\delta B\eta^{\nu\mu},A_{\nu}\right\} =-\int_{\sigma}d\sigma_{\mu}%
\frac{1}{2}\left\{ \delta\pi^{\nu\mu},A_{\nu}\right\} \,.
\]

The induced variations using the same procedure will become%
\begin{equation}
\bar{\delta}A_{\lambda}\left( x\right) =i\int_{\sigma}d\sigma_{\mu}^{\bar {x}%
}\left[ A_{\lambda}\left( x\right) ,A_{\nu}\left( \bar{x}\right)
\right] \delta\pi^{\nu\mu}\left( \bar{x}\right)  \label{varBarA}
\end{equation}%
\begin{equation}
\bar{\delta}\pi^{\alpha\beta}\left( x\right) =i\int_{\sigma}d\sigma_{\mu }^{%
\bar{x}}\left[ \pi^{\alpha\beta}\left( x\right) ,A_{\nu}\left( \bar {x}%
\right) \right] \delta\pi^{\nu\mu}\left( \bar{x}\right) \,.
\label{varBarPi}
\end{equation}

The four equations (\ref{varA}-\ref{varBarPi}) constitute a system
of functional relations. If we take only the $\beta=0$ components,
and impose a kinematical independence between
$\pi^{\nu}=\pi^{\nu0}=F^{\left[ \nu0\right]
}+B\eta^{\nu0}$ and $A_{\nu}$,%
\[
\delta A_{\lambda}\left( x\right)
=-i\int_{\sigma}d\sigma_{0}^{\bar{x}}\left[ A_{\lambda}\left(
x\right) ,\pi^{\nu}\left( \bar{x}\right) \right] \delta
A_{\nu}\left( \bar{x}\right)
\]%
\[
i\int_{\sigma}d\sigma_{0}^{\bar{x}}\left[ A_{\lambda}\left( x\right)
,A_{\nu}\left( \bar{x}\right) \right] \delta\pi^{\nu}\left( \bar
{x}\right) =-i\int_{\sigma}d\sigma_{0}^{\bar{x}}\left[
\pi^{\alpha}\left( x\right) ,\pi^{\nu}\left( \bar{x}\right) \right]
\delta A_{\nu}\left( \bar{x}\right) =0
\]%
\[
\bar{\delta}\pi^{\alpha}\left( x\right) =i\int_{\sigma}d\sigma_{0}^{\bar{x}}%
\left[ \pi^{\alpha}\left( x\right) ,A_{\nu}\left( \bar{x}\right)
\right] \delta\pi^{\nu}\left( \bar{x}\right) \,.
\]

One of the possible solutions of this functional relation system is%
\[
\left[ \pi^{\alpha}\left( x\right) ,\pi^{\nu}\left( \bar{x}\right)
\right] _{0}=\left[ A_{\lambda}\left( x\right) ,A_{\nu}\left( \bar
{x}\right) %
\right] _{0}=0
\]%
\[
\left[ A_{\lambda}\left( x\right) ,\pi^{\nu}\left( \bar{x}\right)
\right] _{0}=i\delta_{\lambda}^{\;\nu}\delta\left( {\bf
x-\bar{x}}\right) \,.
\]

\section{Proca's Field\label{sec3}}

The massive term of the Lagrangian density operator can be written as \cite%
{NakaProca}%
\[
L_{P}=\frac{m^{2}}{2}A_{\mu }A^{\mu }\Longrightarrow \delta L_{P}=\frac{m^{2}%
}{2}\left( \delta A_{\mu }A^{\mu }+A_{\mu }\delta A^{\mu }\right) =\frac{%
m^{2}}{2}\left\{ A^{\mu },\delta A_{\mu }\right\} \,,
\]%
where $m$ is the parameter of mass of Proca's Field.

Note that this variation does not contain a surface term. Then, we
see that
the only equation of motion for the Proca's Field\ that changes is%
\begin{equation}
\partial _{\mu }F^{\left[ \mu \nu \right] }-\partial ^{\nu }B+m^{2}A^{\nu }=0
\label{EqMP}
\end{equation}%
and the generator of functional variations of fields is%
\[
G=\int_{\sigma }d\sigma _{\mu }\left( F^{\left[ \nu \mu \right]
}+B\eta ^{\nu \mu }\right) \delta A_{\nu }\,.
\]

The above generator is the same as we obtained before with the free
electromagnetic field, then the commutation relations for Proca's
Field are also the same.

\section{Dynamical Variables\label{sec4}}

Now we will try to determine the dynamical quantities for both
systems studied before using the analysis of Cauchy's data. This is
a powerful
instrument to separate the dynamical and constrained sectors of a theory %
\cite{DadosCauchy}, making use only of the equations of motion of
the system, and so it can be applied even in the case of a system of
linear field operators, as it was considered here.

\subsection{Proca's Field}

Let us look again the equations of motion of Proca's Field. Substituting (%
\ref{DefF}) in (\ref{EqMP}), we have%
\begin{equation}
\partial _{\mu }\partial ^{\mu }A^{\nu }-\partial _{\mu }\partial ^{\nu
}A^{\mu }-\partial ^{\nu }B+m^{2}A^{\nu }=0  \label{EqMPAmu}
\end{equation}

Now, substituting (\ref{Lorentz}) in the above equation,%
\begin{equation}
\square A^{\nu }+m^{2}A^{\nu }-\left( 1-\alpha \right) \partial
^{\nu }B=0\,. \label{AeB}
\end{equation}

Taking the four-divergence of (\ref{EqMPAmu}) we find%
\begin{equation}
\square B-m^{2}\partial_{\nu}A^{\nu}=0  \label{BeA}
\end{equation}

If we take the gradient of equation (\ref{Lorentz}) and substitute in (\ref%
{EqMPAmu}) multiplied by $\alpha$ we see that%
\begin{equation}
\alpha\left( \square A^{\nu}+m^{2}A^{\nu}\right) +\left(
1-\alpha\right)
\partial^{\nu}\partial_{\mu}A^{\mu}=0  \label{EqCampoAP}
\end{equation}
Multiplying (\ref{Lorentz}) by $m^{2}$ and substituting in (\ref{BeA}),%
\begin{equation}
\square B+\alpha m^{2}B=0  \label{EqCampoBP}
\end{equation}

From this last equation, we can see directly that the auxiliary
scalar field $B$ is a dynamical quantity. On the other hand, the $0$
component of
equation\thinspace (\ref{EqCampoAP}) can be written as%
\[
\partial ^{0}\partial _{0}A^{0}=-\left( 1-\alpha \right) \partial
^{0}\partial _{i}A^{i}-\alpha \left( \partial ^{i}\partial
_{i}A^{0}+m^{2}A^{0}\right)
\]

Once that the above equation involves temporal derivatives of second
order of $A^{0}$, and derivatives of the components $A^{i}$, we need
to check if these last are dynamical quantities.

So, taking the $k$ component of equation\thinspace
(\ref{EqCampoAP}), we
obtain%
\[
\partial ^{0}\partial _{0}A^{k}=-\partial ^{i}\partial _{i}A^{k}-m^{2}A^{k}-%
\frac{\left( 1-\alpha \right) }{\alpha }\partial ^{k}\left( \partial
_{0}A^{0}+\partial _{i}A^{i}\right)
\]%
that show us that, for $\alpha \not=0$, we can determine the
$A^{k}$\ as dynamical quantities.

Since the $A^{k}$\ are established, we may affirm that $A^{0}$ is
also a dynamical quantity. Thus, it becomes clear that Proca's Field
is a system whithout constraints. This result could also be obtained
in a more direct
manner (which includes the case $\alpha =0$) making use of equations (\ref%
{AeB}) and (\ref{EqCampoBP}). Once that (\ref{EqCampoBP}) tell us
that $B$
is a dynamical quantity, we can isolate the second temporal derivative of $%
A^{\nu }$ in (\ref{AeB}):%
\[
\partial ^{0}\partial _{0}A^{\nu }=-\partial ^{i}\partial _{i}A^{\nu
}+\left( 1-\alpha \right) \partial ^{\nu }B-m^{2}A^{\nu }
\]%
and with the knowledge of $B$, $A^{\nu }$, and their first order
temporal derivatives in a given instant, we are able to determine
these fields in any future moment.

\subsection{Eletromagnetic Field}

Following the same steps used in the case of Proca's Field, but using (\ref%
{EqMaxwell}) instead of (\ref{EqMP}), we obtain

\[
\square B=0
\]%
\[
\partial^{0}\partial_{0}A^{0}=\alpha\partial_{i}\left(
\partial^{0}A^{i}-\partial^{i}A^{0}\right) -\partial^{0}\partial_{i}A^{i}
\]

\[
\partial^{0}\partial_{0}A^{k}=-\partial^{i}\partial_{i}A^{k}-\frac{\left(
1-\alpha\right) }{\alpha}\partial^{k}\left(
\partial_{0}A^{0}+\partial _{i}A^{i}\right)
\]

This system of equations show us that the evolution in time of any
component of $A^{\mu }$ is coupled to evolution of the others
components, what is a consequence of the constraint given by the
auxiliary field $B$ - see equation (\ref{Lorentz}). It shows us that
the five quantities $B$, $A^{\nu } $ are dynamical in the case
$\alpha \not=0$.

As we made before in the case of Proca's Field, we can also obtain
the second order temporal derivatives of all components of $A^{\mu
}$ as a
function of derivatives of $B$, directly from the equation%
\[
\square A^{\nu }-\left( 1-\alpha \right) \partial ^{\nu }B=0
\]

So, we see that Nakanishi's formalism gives us a good limit for the
massless field from Proca's equations. We can affirm this once that
all results for the free electromagnetic field may be obtained from
the massive case, making $m=0$.

\section{Final Remarks\label{sec5}}

The union of B-field, Schwinger and Palatini's formalisms have
become more fertile since that an equivalent form of Lorentz
condition of Classical
Electrodynamic can be found as a representative identity between operators (%
\ref{Lorentz}), which implies in a wave equation for the expected
values of the operator $A^{\nu }$ between the physical states
\cite{NakaArt}. We also see that the equations of motion, as well as
the commutation relations for the field operators, can be obtained
in a covariant form from an unique variational principle, being, in
that way, automatically self-consistent.

Another advantage is that this procedure may be generalized in a
natural manner by the inclusion of source terms or spinorial fields
in the Lagrangian density operator, in order to construct a theory
of Quantum Electrodynamic in a general linear covariant gauge.

The analysis of the constraints of the theory can be implemented \
through {\em Cauchy's Data} \cite{DadosCauchy}, and indicate that
the B-field formalism provide a good limit for the massless
vectorial field from the Proca's field \cite{NakaProca}.

In future perspectives, we intend to implement the same method to
make a preliminar study of quantized gravitational field.\bigskip

\section*{Acknowledgements}
%We would like to thank ...........
The authors acknowledge the suggestions and comments of R. G.
Teixeira. B.M.P. would like to thank CNPq and FAPESP (grant
02/00222-9) for partial financial support. C. A. M. M. thanks FAPESP
(grants 99/09091-0 and 01/12584-0) for full support. P. J. P. thanks
CNPq for partial support and the CTA staff for the incentive.

%\appendix
%\section{First Appendix} %Empty argument \section{} yields `Appendix'.
%
%\section{Second Appendix}

\end{document}